\documentclass[conference]{IEEEtran}
\IEEEoverridecommandlockouts
\usepackage{cite}
\usepackage{amsmath,amssymb,amsfonts}
\usepackage{algorithmic}
\usepackage{graphicx}
\usepackage{textcomp}
\usepackage{xcolor}
\usepackage{booktabs}
\usepackage{tikz}
\usetikzlibrary{arrows.meta, positioning, fit, backgrounds}
\graphicspath{{figures/}}

\def\BibTeX{{\rm B\kern-.05em{\sc i\kern-.025em b}\kern-.08em
    T\kern-.1667em\lower.7ex\hbox{E}\kern-.125emX}}

\begin{document}

\title{Weather Data Spoofing Attacks on Rain-Adaptive Millimeter-Wave Frequency Selection in V2X Communication Networks}

\author{\IEEEauthorblockN{1\textsuperscript{st} Rasheed Bello}
\IEEEauthorblockA{\textit{Department of Engineering} \\
\textit{South Carolina State University}\\
Orangeburg, SC USA \\
rbello@scsu.edu}
\and
\IEEEauthorblockN{2\textsuperscript{nd} Idreez Yusuf}
\IEEEauthorblockA{\textit{Department of Engineering} \\
\textit{South Carolina State University}\\
Orangeburg, SC USA \\
iyusuf@scsu.edu}
\and
\IEEEauthorblockN{3\textsuperscript{rd} Justice Adjei Owusu}
\IEEEauthorblockA{\textit{Computational Data Science and Eng.} \\
\textit{North Carolina A\&T State University}\\
Greensboro, NC USA \\
jadjeiowusu@aggies.ncat.edu}
\and
\IEEEauthorblockN{4\textsuperscript{th} Oluwatobiloba Aiyewunmi}
\IEEEauthorblockA{\textit{Computational Data Science and Eng.} \\
\textit{North Carolina A\&T State University}\\
Greensboro, NC USA \\
opaiyewunmi@aggies.ncat.edu}
\and
\IEEEauthorblockN{5\textsuperscript{th} Gurcan Comert}
\IEEEauthorblockA{\textit{Computational Data Science and Eng.} \\
\textit{North Carolina A\&T State University}\\
Greensboro, NC USA \\
gcomert@ncat.edu}
\and
\IEEEauthorblockN{6\textsuperscript{th} Judith Mwakalonge}
\IEEEauthorblockA{\textit{Department of Engineering} \\
\textit{South Carolina State University}\\
Orangeburg, SC USA \\
jmwakalo@scsu.edu}
\and
\IEEEauthorblockN{7\textsuperscript{th} Esmail Abuhdima}
\IEEEauthorblockA{\textit{ Engineering Technology }\\
\textit{ECPI University}\\
Newport News, VA USA \\
eabuhdima@ecpi.edu}
\and
\IEEEauthorblockN{8\textsuperscript{th} Abdulmajid Mrebit}
\IEEEauthorblockA{\textit{Engineering and Computer Science} \\
\textit{Benedict College}\\
Columbia, SC USA \\
a.mrebit@benedict.edu}
\and
\IEEEauthorblockN{9\textsuperscript{th} Rajab Ataai}
\IEEEauthorblockA{\textit{Engineering and Computer Science} \\
\textit{Benedict college}\\
Columbia, SC USA \\
Rajab.Ataai@benedict.edu}
\and
\IEEEauthorblockN{10\textsuperscript{th} Vaidyan Varghese}
\IEEEauthorblockA{\textit{College of Computer \& Cyber Sciences} \\
\textit{Dakota State University}\\
Madison, SD USA \\
varghese.vaidyan@dsu.edu}
}

\maketitle

\begin{abstract} 
Connected vehicles use millimeter-wave (mmWave) sidelinks for the data rates cooperative driving demands, and emerging designs select the carrier band from sensed rainfall. We show that this weather awareness is an attack surface: an adversary who spoofs only the rainfall input dictates the victim's carrier frequency, and through it its communication range, without transmitting on the channel. We evaluate the attack in MilliCar, an ns-3 module that runs the selected band as the real 3GPP NR V2X carrier with per-band propagation, beamforming, and blockage. Forcing the band up to 73 GHz holds an eight-vehicle platoon's reliable range at 38 m while the honest baseline doubles it to 82 m; forcing it down to 5 GHz sustains 97\% long-range reception but collapses the transport block to a third and quadruples long-range latency to 12.5 ms. We then implement the defense the mechanism implies. Rain loss grows linearly with distance while path loss grows logarithmically, so a receiver that tests measured SINR against the attenuation its reported weather predicts flags force-up with 98\% probability within 1.5 s at a 1\% false-alarm rate, and re-selection then restores long-range reception from 60\% to 75\%. The same test is structurally blind to force-down, because the 5 GHz fallback is nearly rain-immune. An advecting rain cell that swings the local rate from 15 to 81 mm/h leaves every result unchanged. Weather-aware band selection therefore requires an authenticated meteorological input; physical cross-checking covers one half of the threat.
\end{abstract}


\section{Introduction}
Connected and automated vehicles (CAVs) exchange awareness, perception and control data under latency constraints that exceed sub-6 GHz throughput. Millimeter-wave (mmWave) bands are the only practical spectrum for gigabit-per-second raw-sensor exchange \cite{choi2016,jameel2019}.

That capacity costs link robustness. mmWave performance varies with blockage, geometry, beam alignment and atmospheric loss \cite{jameel2019,wang2018}, so multi-band designs treat the band as a control variable \cite{collperales2019,zhou2016}. Rain attenuation rises steeply with frequency \cite{peric2017,huang2019} under ITU-R P.838 \cite{itu_p838}, so a weather-aware stack steps down when sensed rainfall erodes its margin. The controller studied here selects among 28, 39, 60 and 73 GHz with a 5 GHz fallback.

Adaptive selection introduces a dependency absent from static designs. A vehicle acting on an external weather report makes that report a control input to its radio, and an input trusted without authentication is one an adversary can supply. Prior work does not examine the case where the falsified value is the weather and the actuator the carrier.

This paper examines that case in MilliCar \cite{drago2020}, which runs the selected band as the actual carrier over a 3GPP V2V channel with beamforming and blockage \cite{mezzavilla2018}, so a forced band change moves real propagation rather than a lookup table. We contribute:
\begin{itemize}
    \item We formalize weather spoofing against rain-adaptive band selection as two attack classes: \textit{force-up} (Type 1) toward higher, rain-sensitive bands and \textit{force-down} (Type 2) toward lower-capacity bands, and evaluate both in the native mmWave channel.
    \item We separate propagation reach from bandwidth cost by running each case under fixed and per-band bandwidth regimes, exposing a trade-off that fixed-bandwidth evaluation hides.
    \item We trace both effects to per-link signal-to-interference-plus-noise ratio (SINR), establishing the reach attack as a propagation effect rather than a consequence of the bandwidth assumption.
    \item We implement a receiver-side physical-consistency detector that exploits the distance scaling of rain loss. It detects force-up within seconds and is structurally blind to force-down, which bounds what any physical defense achieves.
\end{itemize}
\section{Related Work}
\begin{figure*}[t]
\centerline{\includegraphics[width=\textwidth]{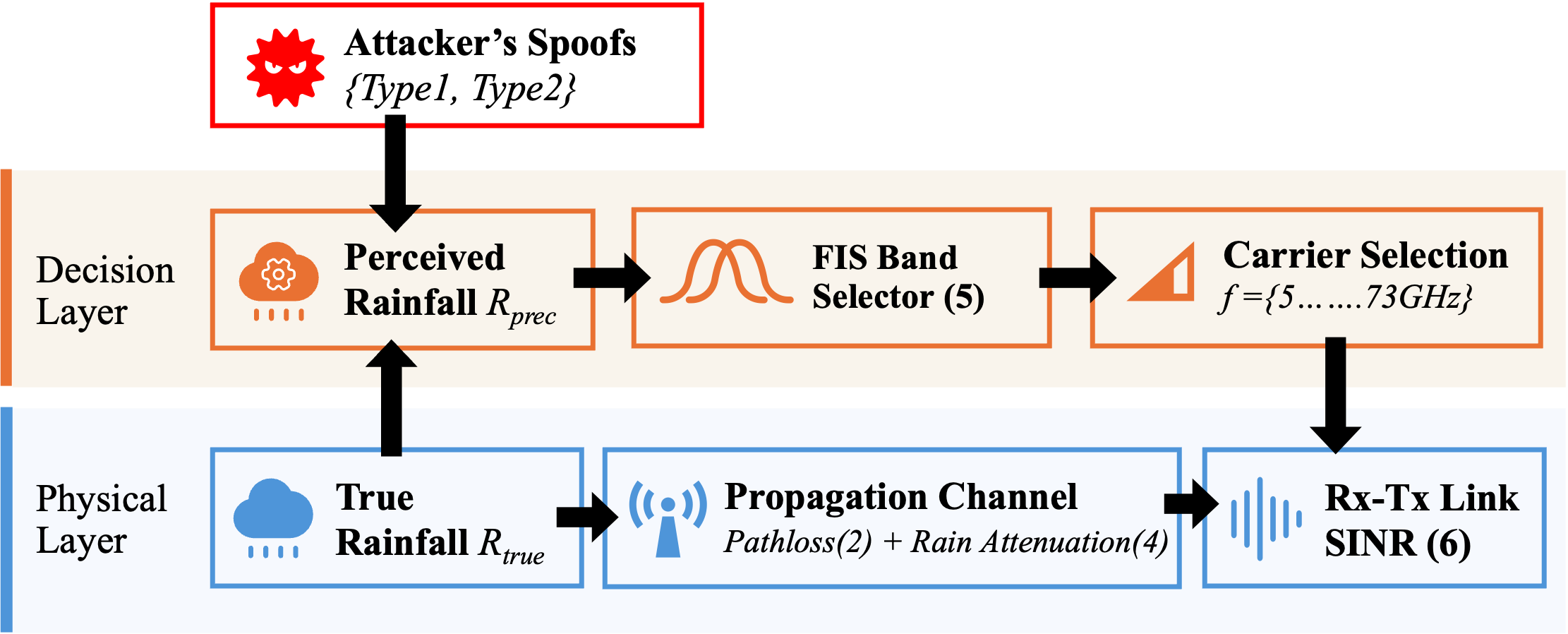}}
\caption{Rain-adaptive band selection and the spoofing path: a falsified rain value overrides perceived rain and forces the carrier, while true rain still attenuates the chosen band.}
\label{fig:attack-logic}
\end{figure*}

The literature motivating mmWave V2X also exposes its dependence on adaptation. Sensing workloads require mmWave bandwidth \cite{choi2016}, propagation surveys name weather as an open problem \cite{jameel2019}, and line-of-sight alignment governs urban coverage \cite{wang2018}. Each treats the signal driving adaptation as honest.

Rain justifies that adaptation and its modeling is mature. Rain attenuation is dynamic rather than a static offset \cite{peric2017}, and Recommendation P.838 standardizes it as a frequency-dependent function of rain rate \cite{itu_p838}. Huang \textit{et al.} validate that law at 26 and 77 GHz and report the long-path distance factor inappropriate for short links, which matters at platoon ranges \cite{huang2019}. Turning the physics into a control loop is equally well studied: Zhou \textit{et al.} select bands from a trusted geolocation database \cite{zhou2016}, and Coll-Perales \textit{et al.} decouple a sub-6 GHz control plane from a mmWave data plane \cite{collperales2019}. Both assume authentic selection inputs. We adopt packet delivery against distance \cite{gonzalezmartin2019} as the reliability metric, resolved per directed link.

Vehicular security supplies the perspective these works omit. Spoofing is a primary threat, and communication-layer attacks reach the control layer \cite{elrewini2020}. Biroon \textit{et al.} detect false data in a cooperative-adaptive-cruise-control platoon with an observer tuned to vehicle dynamics \cite{biroon2022}; that observer watches the control state, so an attack on the radio input passes beneath it. Baee \textit{et al.} quantify the latency IEEE 1609.2 adds in dense traffic \cite{baee2019}, which explains why low-stakes beacons are left unsigned, and Rasheed \textit{et al.} secure the mmWave data path but not the weather input \cite{rasheed2021}. None measures spoofed weather steering native band selection.

\section{Threat Model and Weather-Aware Selection}
\subsection{Adaptive selection and its inputs}
The vehicle chooses a carrier from the band set
\begin{equation}
\mathcal{F} = \{5,\ 28,\ 39,\ 60,\ 73\}\ \text{GHz},
\label{eq:bands}
\end{equation}
where the four mmWave bands carry high-rate traffic and 5 GHz is the fallback. Higher bands offer wider channels but larger loss, since the line-of-sight path loss of the 3GPP urban V2V model grows with carrier frequency,
\begin{equation}
\mathrm{PL}(d,f) = 38.77 + 16.7\log_{10} d + 18.2\log_{10} f,
\label{eq:pathloss}
\end{equation}
with $\mathrm{PL}$ the median in dB, $d$ the three-dimensional separation in metres and $f$ the carrier in GHz, about which $\sigma_{\mathrm{SF}}=3$ dB of log-normal shadowing applies \cite{3gpp37885,drago2020}. The simulator also applies the non-line-of-sight branch $36.85+30\log_{10}d+18.9\log_{10}f$ and an additional log-normal vehicle-blockage loss, selected per link by the V2V-Urban condition model.

Rain adds a frequency-dependent term. Following ITU-R P.838 \cite{itu_p838},
\begin{equation}
\gamma(f,R) = k_f\, R^{\alpha_f}\quad[\text{dB/km}],
\label{eq:itu}
\end{equation}
with $R$ in mm/h. For the calibrated dataset used here, $(k_f,\alpha_f)$ is $(0.00022,1.694)$ at 5 GHz and $(0.205,0.968)$, $(0.422,0.874)$, $(0.861,0.766)$, $(1.076,0.727)$ at 28, 39, 60, and 73 GHz; $\alpha_f$ falls toward $0.73$ as frequency rises, as P.838 predicts. Over a link of length $d$ in metres the excess loss is
\begin{equation}
A_{\mathrm{rain}}(f,R,d) = \gamma(f,R)\,\frac{d}{1000}\quad[\text{dB}].
\label{eq:rainloss}
\end{equation}
At 50 mm/h this gives 0.17 dB/km at 5 GHz against 9.04, 12.89, 17.20 and 18.49 dB/km at 28, 39, 60 and 73 GHz. The two-order-of-magnitude gap makes the fallback attractive to an attacker and, as Section~\ref{sec:defense} shows, makes the lie about it unobservable. The controller takes the highest band whose budget stays within a threshold,
\begin{equation}
f^{\ast} = \max\big\{ f \in \mathcal{F}_{\mathrm{mm}} : A_0(f) + \mathcal{L}_f(R_{\mathrm{perc}}) \leq A_{\mathrm{th}} \big\},
\label{eq:selector}
\end{equation}
over $\mathcal{F}_{\mathrm{mm}}=\{28,39,60,73\}$ GHz, falling back to 5 GHz when the set is empty. $R_{\mathrm{perc}}$ is the perceived rain rate, $A_0(f)$ the zero-rain budget (65.35, 68.49, 69.93 dB at 39, 60, 73 GHz), and $\mathcal{L}_f$ is \eqref{eq:rainloss} at the 1 km design reference of the fuzzy inference system \cite{abuhdima2024fis,abuhdima2026fis}, interpolated from a 20-point table. $A_0(f)$ is that controller's design-time budget on its own link-budget convention rather than \eqref{eq:pathloss}, and only its ordering across bands drives selection. With $A_{\mathrm{th}}=74$ dB, 73 GHz holds to about 6 mm/h, then 60 and 39 GHz, reaching 28 GHz at 32 mm/h. We implement 28 GHz as an unconditional floor, so the honest arm never reaches the fallback and only a spoofed report does.

\subsection{Adversary and the two attack types}
The attack exploits a separation between spoofable perception and unspoofable propagation (Fig.~\ref{fig:attack-logic}). The adversary cannot alter the rain field, the trajectories or the radio; it supplies a false value through a compromised or spoofed beacon that the controller ingests as $R_{\mathrm{perc}}$, which is feasible whenever the weather channel is unauthenticated \cite{elrewini2020}. Equations \eqref{eq:pathloss}--\eqref{eq:rainloss} keep acting on the true rate $R_{\mathrm{true}}$; only the selector \eqref{eq:selector} responds to the spoofed value. \emph{Attack Type 1} reports clear sky, pinning selection at 73 GHz regardless of the real downpour, so the victim operates at the worst-propagating band while true rain attenuates it. \emph{Attack Type 2} reports a deluge, driving the controller to the 5 GHz fallback, so the victim gains reach but abandons the bandwidth that motivates mmWave operation. One denies reach, the other capacity.

\section{Native mmWave Evaluation}
\subsection{Platform and channel}
MilliCar \cite{drago2020} implements the 3GPP NR V2X sidelink PHY and MAC over the spatial mmWave channel of \cite{mezzavilla2018}, instantiating the selected band as the actual carrier, so \eqref{eq:pathloss}, beamforming from a uniform planar array with DFT codebook and the LOS/NLOS/NLOSv states of the V2V-Urban scenario follow from the chosen frequency. The channel carries no rain term, so we chain \eqref{eq:rainloss} onto it as an additional per-link loss.

We model rain two ways. The primary arm holds one rate over the platoon, isolating the attack mechanism from any second varying factor. The second uses an advecting rain cell: the exponential EXCELL profile $R(p)=R_M\exp(-\lVert p-c\rVert/\rho_0)$ \cite{capsoni1987} translating at 10 m/s under the frozen-field assumption of the synthetic storm technique \cite{matricciani1996sst}. Attenuation becomes the path integral of $\gamma(f,R(s,t))$ along each link, varying per link and per instant. The cell passes over the platoon mid-run, and $R_M$ is calibrated so platoon-mean rain equals the nominal value, attributing any difference to variability rather than to a different quantity of water. Since $\rho_0$ is exponentially distributed about a mean fixed by peak rate, we sweep it as a multiple $\bar{\rho}_0$ of that mean at 20, 37.5 and 50 mm/h under fixed width.

The platoon holds eight vehicles, the largest collision-free group under numerology 3. Because the study measures reach, we treat the geometry as an instrument rather than traffic: vehicles travel at 20 m/s with fixed spacing, holding every directed-link distance constant and placing 21 distinct distances between 15 and 290 m, so one 30 s run samples the whole reception-versus-distance curve at known distances. A car-following trace would collapse the long links that carry the result. Each vehicle sends a 200-byte awareness message to every other at 10 Hz, giving 56 directed links.

\begin{figure*}[t]
\centerline{\includegraphics[width=\textwidth]{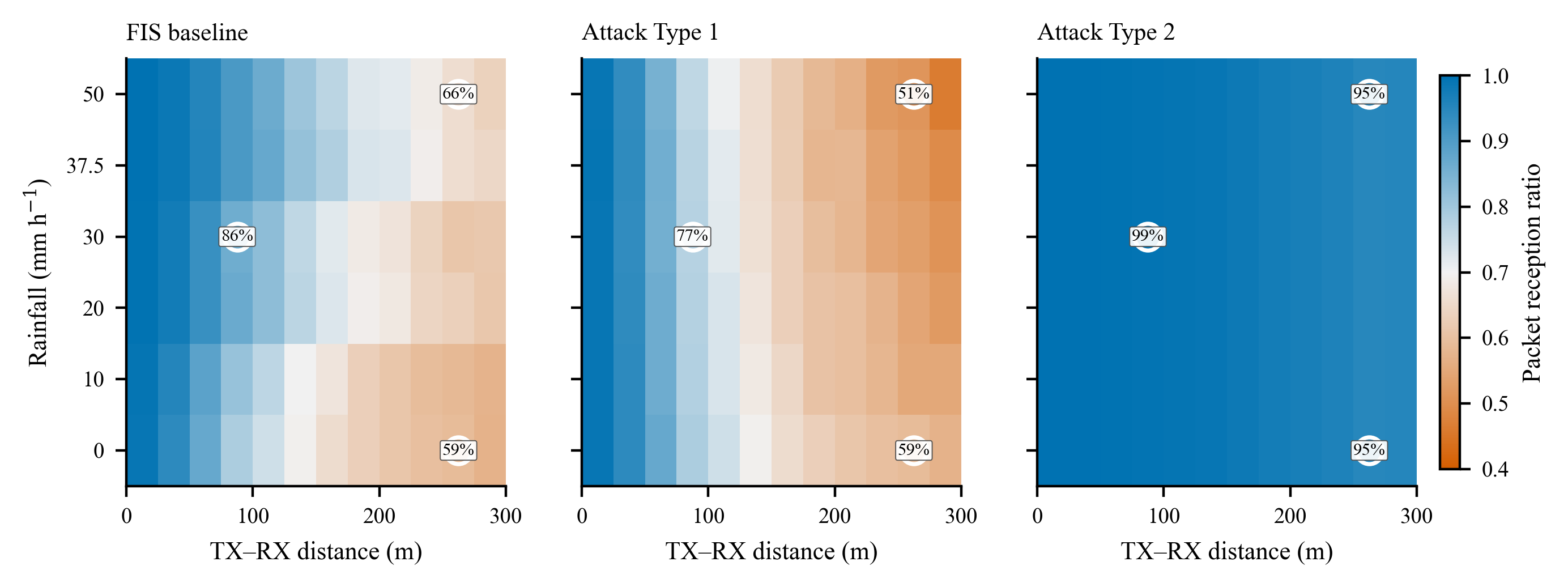}}
\caption{Packet reception ratio over link distance and true rainfall, per scenario, fixed-width regime (ten-seed mean). The real-width regime is near-identical in delivery, as Section~\ref{sec:regimes} shows.}
\label{fig:heatmap}
\end{figure*}

\subsection{Two bandwidth regimes}
Bandwidth is the second lever a band switch moves, so we run every configuration twice. The \emph{fixed-width} regime gives all bands 100 MHz, holding capacity constant so only carrier frequency differs; the \emph{real-width} regime assigns deployable widths, 20 MHz at 5 GHz and 100, 200 and 400 MHz above. Bandwidth sets the noise floor and transport-block budget, not the path loss:
\begin{equation}
\mathrm{SINR} = \frac{P_{\mathrm{tx}}\, G_{\mathrm{bf}}\, 10^{-A_{\mathrm{tot}}/10}}{k\,T_0\,B\,F + I},
\label{eq:sinr}
\end{equation}
where $A_{\mathrm{tot}} = \mathrm{PL}(d,f^{\ast}) + A_{\mathrm{rain}}(f^{\ast},R_{\mathrm{true}},d)$ is the total loss in dB, $P_{\mathrm{tx}}=30$ dBm is the transmit power, $G_{\mathrm{bf}}$ is the combined transmit and receive beamforming gain of the uniform planar arrays under perfect alignment, $B$ is the channel bandwidth in Hz, $F=5$ dB is the noise figure, $k$ is Boltzmann's constant, and $T_0=290$ K. The interference term $I$ is negligible by construction: one vehicle transmits per slot, so the group is collision-free and the denominator is thermal-noise dominated, which attributes the reach results to propagation alone. Widening $B$ raises the noise floor, so the real-width regime moves both capacity and the SINR deciding whether a long link closes.

\subsection{Parameters and metrics}
Table~\ref{tab:params} lists the configuration. We sweep rainfall over $\{0,10,20,30,37.5,50\}$ mm/h with ten channel realizations per case, reporting mean and standard deviation across seeds. The primary metric is the packet reception ratio (PRR) of a directed link, the count received divided by the count sent, resolved against link distance and aggregated beyond 100 m. Reliable range is the centre of the farthest 25 m bin whose PRR stays at or above 90\%. The physical layer reports the SINR of \eqref{eq:sinr} for every decoded transport block, and transport-block size and latency expose capacity.

\begin{table}[t]
\caption{Simulation parameters}
\label{tab:params}
\centering
\renewcommand{\arraystretch}{1.15}
\begin{tabular}{@{}ll@{}}
\toprule
\textbf{Parameter} & \textbf{Value} \\
\midrule
Simulator, channel & ns-3.36.1 MilliCar \cite{drago2020}, 3GPP V2V-Urban \\
Numerology & 3 (120 kHz SCS, 8 slots/subframe) \\
Platoon & 8 vehicles, 15--290 m spread, 20 m/s \\
Antenna, beamforming & uniform planar array, DFT codebook \\
Tx power, noise fig. & 30 dBm, 5 dB ($T_0\!=\!290$ K) \\
Carrier bands & 5, 28, 39, 60, 73 GHz \\
Bandwidth (fix / real) & 100 / \{20,100,200,400\} MHz \\
Channel update & 100 ms \\
CAM traffic & 200 B at 10 Hz, full mesh, UDP \\
RLC, adaptation & UM 500 kB, AMC (MCS 0--28) \\
Switching budget & $A_{\mathrm{th}} = 74$ dB \\
Rain, seeds, time & 0--50 mm/h (6 pts), 10 seeds, 30 s \\
PRR binning, long range & 25 m bins, $>100$ m \\
Detector window, alarm & $>150$ m links, 1\% false-alarm rate \\
\bottomrule
\end{tabular}
\end{table}

\section{Results and Discussion}
\subsection{Forcing the band up is a reach attack that scales with rain}
Fig.~\ref{fig:heatmap} maps PRR across distance and rainfall. The baseline shows the protective behavior of adaptation: at fixed distance its reception improves as rain rises, because the controller steps down to a better-propagating band. Attack Type 2 saturates the map, since 5 GHz barely feels the rain. Attack Type 1 inverts the baseline, decaying toward long range and staying low as rain intensifies, because the victim is pinned at 73 GHz, the band with least margin for the added loss.

\begin{figure}[!t]
\centerline{\includegraphics[width=\columnwidth]{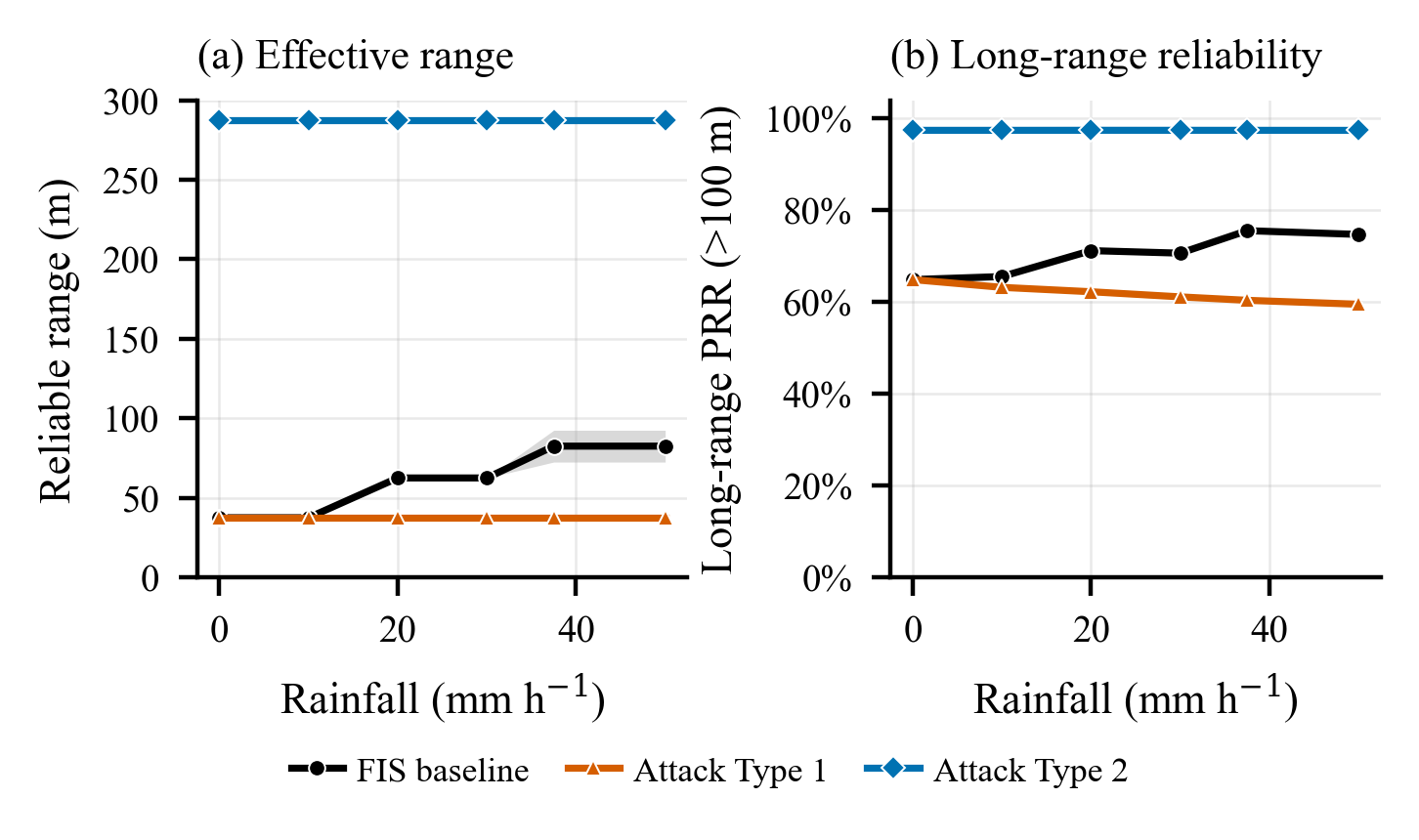}}
\caption{Reach vs.\ rainfall, fixed-width regime (ten-seed mean$\pm$std). (a) Reliable range (PRR $\geq 90\%$). (b) Long-range PRR ($>100$ m).}
\label{fig:reach}
\end{figure}

Fig.~\ref{fig:reach} quantifies it. Honest beacons more than double the reliable range, 38 m in light rain to 82 m at 50 mm/h. Attack Type 1 erases that gain, holding 38 m across the sweep while its long-range PRR drifts from 65\% to 60\% and the baseline climbs from 65\% to 75\%. Attack Type 2 holds 288 m and 97\%, the sub-6 GHz carrier being effectively rain-immune.

\subsection{SINR explains the reach, and the result survives the bandwidth assumption}
\label{sec:regimes}
The reach outcomes follow from link budget. Mean SINR beyond 200 m at 50 mm/h is 35.6 dB for the 5 GHz carrier of Attack Type 2, 19.7 dB for the baseline and 8.6 dB for the 73 GHz carrier of Attack Type 1, whose long links therefore fall below the decoding margin first. The real-width regime widens this separation: by \eqref{eq:sinr} the 400 MHz channel at 73 GHz strips 5.2 dB from Attack Type 1, to 3.5 dB, while the 20 MHz fallback channel lifts its budget by 6.5 dB.

The two levers partly cancel in delivery: the wider channel costs those 5.2 dB but returns capacity, so long-range PRR for Attack Type 1 moves only from 59.5\% to 60.0\% between regimes and the baseline is unchanged. Realistic bandwidth therefore establishes the reach attack as a propagation effect and rules out constant bandwidth as its cause. For Attack Type 2 it instead exposes a capacity harm that fixed width conceals.

\subsection{Forcing the band down trades reach for capacity}
\label{sec:cost}
Constant bandwidth hides what the sub-6 GHz fallback surrenders. Under fixed width every band carries a comparable transport block on long links; under real width the 20 MHz fallback collapses from 231 to 79 bytes, a third of the capacity the mmWave bands sustain. The narrowing also queues the cooperative messages, raising long-range latency for Attack Type 2 from 0.8 to 12.5 ms. Forcing the band down therefore trades throughput and timeliness for rain-proof reach, mirroring the reach denial of Attack Type 1.

\subsection{A physical-consistency detector and its blind spot}
\label{sec:defense}
An unauthenticated weather report is therefore a direct actuator of the physical layer. Authentication is the direct answer, but it adds latency under load \cite{baee2019}, and detectors watching the control state would not flag a plausible clear-sky report \cite{biroon2022}. We quantify how much of the threat a receiver eliminates from measurements it already collects.

Rain loss is linear in $d$ by \eqref{eq:rainloss} while path loss is logarithmic by \eqref{eq:pathloss}, so an unmodelled rain term leaves a distinctive ramp in observed SINR. A vehicle holding a per-band clear-sky profile $S_{\mathrm{ref}}(f,d)$, calibrated in dry operation, expects
\begin{equation}
S_{\mathrm{exp}}(f,d) = S_{\mathrm{ref}}(f,d) - \gamma(f,R_{\mathrm{perc}})\,\frac{d}{1000}
\label{eq:expected}
\end{equation}
under the \emph{reported} rain rate, and forms the residual $\Delta=\mathrm{SINR}_{\mathrm{obs}}-S_{\mathrm{exp}}$ per decoded transport block. Honest operation leaves $\Delta$ slightly negative, drifting from $-0.5$ to $-1.1$ dB as rain rises, so the null must pool every rain rate; a force-up spoof reporting $R_{\mathrm{perc}}=0$ during real rain leaves $\Delta=-\gamma(f^{\ast},R_{\mathrm{true}})d/1000$. The detector averages $\Delta$ over $W$ blocks beyond 150 m and alarms below a threshold set for a 1\% false-alarm rate against that composite null. We calibrate $S_{\mathrm{ref}}$ on five channel realizations and report on the five held out.

\begin{figure}[t]
\centerline{\includegraphics[width=\columnwidth]{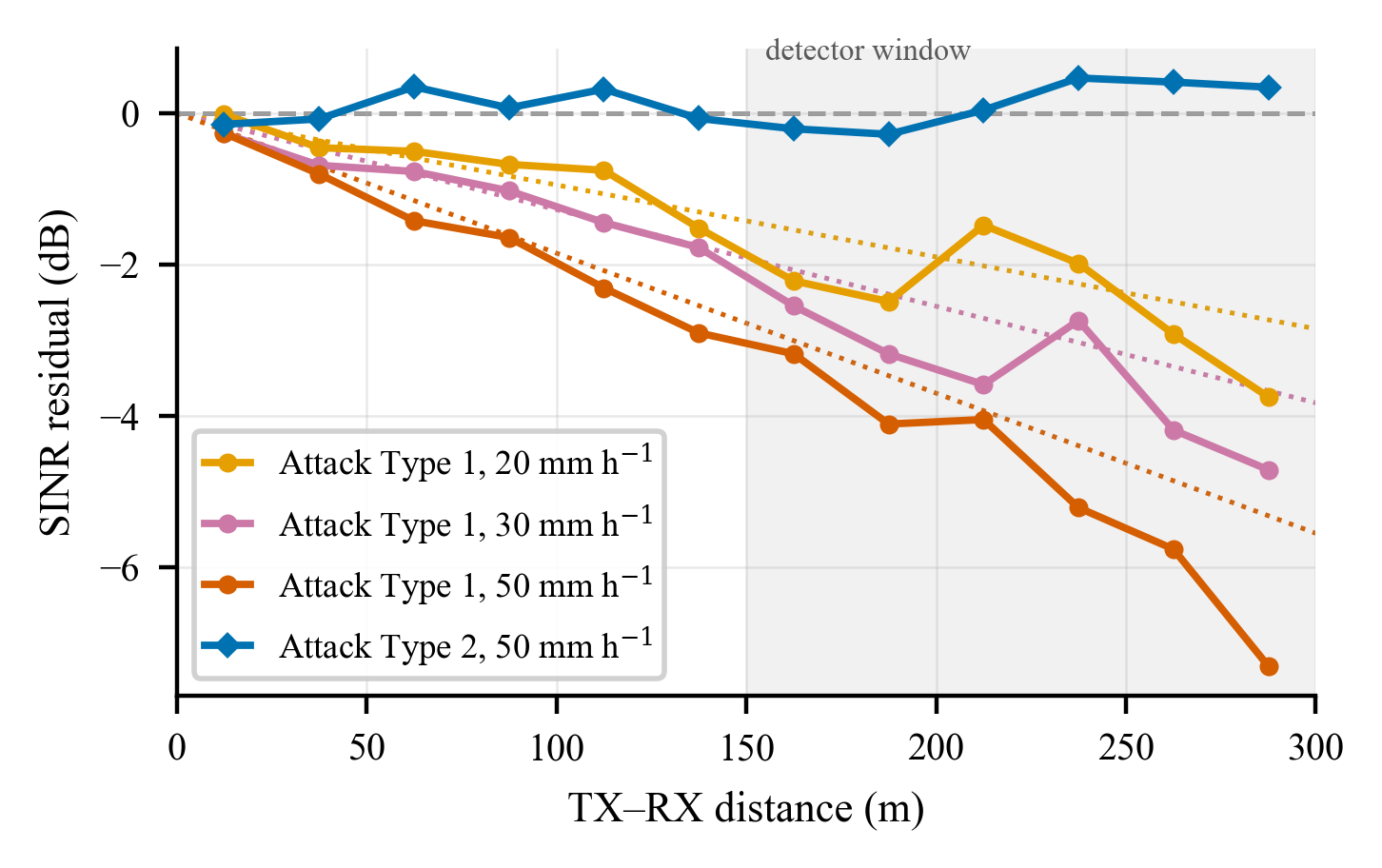}}
\caption{SINR residual against the attenuation the reported weather predicts; dotted lines are the ITU-R expectation.}
\label{fig:defense}
\end{figure}

Fig.~\ref{fig:defense} shows the signature: the residual reaches $-5$ dB on average beyond 150 m, falling below the ITU-R prediction at the longest links, while moving under 0.3 dB at 15 m, which is why the detector watches long links. At 50 mm/h it reaches 68\% within 20 blocks and 98\% within 50, or 0.6 and 1.5 s at the measured 33.5 blocks per second per vehicle; at 30 mm/h it needs 100 blocks, 3.0 s, for 96\%. Detection falls to 19\% at 10 mm/h, but the attack weakens for the same reason, both scaling with $\gamma(f^{\ast},R_{\mathrm{true}})$.

Detection enables mitigation: the victim discards the flagged report and selects from its own link budget, which is the baseline arm of the sweep, so the benefit is measured rather than assumed. Re-selection lifts long-range PRR from 59.5\% to 74.7\% and reliable range from 38 to 82 m, recovering the entire attack effect.

The test is structurally blind to Attack Type 2. Since $\gamma(5,50)=0.17$ dB/km, a reported deluge and a reported clear sky differ by 0.05 dB over a 290 m link, and the measured residual is flat at $+0.07$ dB. A better statistic cannot close this gap: the rain immunity that makes the fallback attractive also makes the lie about it unobservable, so no receiver-side test on a rain-immune carrier detects it. Physical cross-checking and input authentication therefore address different halves of the threat, and the force-down half justifies the IEEE~1609.2 cost \cite{baee2019} on the weather channel.

The detector uses the same ITU-R model the simulator injects, so these rates bound what a deployment facing model mismatch achieves. The 7.5 dB per-block spread from blockage and LOS/NLOSv transitions is not matched and sets the window length.

\subsection{Robustness to a varying rain field}
\label{sec:rainfield}
Table~\ref{tab:rainfield} reports the cell arm. Shrinking the cell drives the local rain rate through a 5.3-fold swing within a single run, yet long-range PRR and detection rate both stay at their uniform-rain values. Scale explains this. Rain contributes only 5.36 dB over a 290 m link at a steady 50 mm/h, so even the 5.3-fold swing moves the attenuation on that link by 4.22 dB, against a 7.5 dB per-block spread from blockage and LOS/NLOSv transitions. A mean-preserving perturbation smaller than the channel's own variability, averaged over thousands of transport blocks, does not move a delivery statistic. The reach attack and its detector therefore depend on the link budget, not on the assumption of uniform rain.

\begin{table}[t]
\caption{Cell arm at 50 mm/h mean rain. Cell size is a multiple of the Capsoni mean $\bar{\rho}_0$; smaller is more variable.}
\label{tab:rainfield}
\centering
\renewcommand{\arraystretch}{1.15}
\begin{tabular}{@{}lcccc@{}}
\toprule
Cell size & Rain swing & $A_{\mathrm{rain}}$ swing & PRR & Detect \\
 & (mm/h) & (dB, 290 m) & ($>$100 m) & ($W\!=\!50$) \\
\midrule
uniform          & 50.0 flat & 0.00 & 59.5\% & 98\% \\
$2.0\bar{\rho}_0$  & 44.8--52.9 & 0.50 & 59.9\% & 93\% \\
$1.0\bar{\rho}_0$  & 39.7--56.1 & 1.01 & 59.5\% & 98\% \\
$0.5\bar{\rho}_0$  & 30.3--63.2 & 2.06 & 59.3\% & 97\% \\
$0.25\bar{\rho}_0$ & 15.3--80.8 & 4.22 & 59.6\% & 96\% \\
\bottomrule
\end{tabular}
\end{table}

\subsection{Limitations}
\label{sec:limits}
Numerology 3 caps the platoon at eight collision-free vehicles, so contention, which would make $I$ in \eqref{eq:sinr} non-negligible, is out of scope. The 5~GHz fallback runs on a PHY tuned for mmWave beamforming, so its absolute reception is an upper bound rather than a calibrated 802.11p result. The cell arm fixes the carrier for a run, so it varies the \emph{propagation} a vehicle experiences but not the \emph{control} decision through which the attack acts; scored against platoon-mean rain the honest controller holds 28 GHz at $0.5\bar{\rho}_0$ and above, as simulated, and would switch twice at $0.25\bar{\rho}_0$, so the model represents a controller whose update period is long relative to a cell passage. Rain stays uncoupled from blockage geometry, a sharp-edged front is a different geometry from a cell, the fixed distance ladder makes the detector's reference profile cleaner than one learned under real mobility, and 73 GHz extrapolates the V2V path-loss table 1.2 dB above its 63 GHz evaluation point.

\section{Conclusion}
Weather awareness becomes a vulnerability when its input is trusted without proof. In a simulator that runs the chosen band as the real carrier, an adversary who spoofs only the rainfall estimate controls the victim's carrier and its range: forcing the band up freezes reliable range at 38 m while the honest baseline doubles it to 82 m, and forcing it down secures reach but collapses capacity to a third and quadruples latency. Per-link SINR accounts for both.

The same mechanism supplies a defense and bounds it. Because rain loss scales linearly with distance where path loss scales logarithmically, a receiver testing measured SINR against the attenuation its reported weather predicts flags force-up within seconds and recovers the full reach loss by re-selecting, without transmitting or trusting an external party. Force-down defeats that test, the fallback being nearly rain-immune, so an authenticated meteorological channel remains necessary for the half of the threat measurement cannot observe. Contention-limited operation, a calibrated sub-6~GHz fallback PHY, within-run re-selection, and authenticated beacons costed against IEEE~1609.2 are the natural extensions.

\section*{Acknowledgment}
This research was partly funded by the U.S. Department of Education, United States, through the HBCU Master’s Program Grant (Grant No. P382G230015), the National Science Foundation, United States, under Grant Nos. 2131080, 2242812, 2200457, 2234920, and 2305470. The authors also gratefully acknowledge the support and constructive feedback provided by advisors, the Yale University ASCEND program, colleagues, and reviewers.

\bibliographystyle{IEEEtran}
\bibliography{Sources_V2}

\end{document}